\documentclass[journal]{IEEEtran}

\usepackage{mathrsfs}
\usepackage{bbm}
\usepackage{amssymb}
\usepackage{bbding}
\usepackage{threeparttable}

\usepackage[mathcal]{euscript}

\usepackage{psfrag,calc,url,bm}
\usepackage{diagbox}
\usepackage{cite}

\usepackage{graphicx}
\usepackage{siunitx}
\usepackage{psfrag}

\usepackage{subfigure}
\usepackage{hyperref}
\usepackage{url}

\usepackage{stfloats}

\usepackage{amsmath}

\usepackage{float}

\usepackage{algorithmic}

\usepackage[ruled,linesnumbered,vlined]{algorithm2e}

\usepackage{color}

\usepackage{boxedminipage}

\usepackage{amsthm}

\usepackage{multirow}

\usepackage{setspace}

\usepackage{soul}
\usepackage{epstopdf}

\usepackage{booktabs}
\allowdisplaybreaks[3]
\begin{document}

\title{AI-Empowered RIS-Assisted Networks: CV-Enabled RIS Selection and  DNN-Enabled Transmission}
\author{Conggang Hu, Yang Lu,~\IEEEmembership{Member,~IEEE}, Hongyang Du,~\IEEEmembership{Graduate Student Member},~Mi Yang~\IEEEmembership{Member,~IEEE}, \\Bo Ai,~\IEEEmembership{Fellow,~IEEE}, Dusit Niyato,~\IEEEmembership{Fellow,~IEEE}
\thanks{Conggang Hu and Yang Lu are with the School of Computer Science and Technology, Beijing Jiaotong University, Beijing 100044, China (e-mail: 23120352@bjtu.edu.cn,yanglu@bjtu.edu.cn).}
\thanks{Hongyang Du and Dusit Niyato are with the School of Computer Science and Engineering, Nanyang Technological University, Singapore 639798 (e-mail: hongyang001@e.ntu.edu.sg,dniyato@ntu.edu.sg).}
\thanks{Mi Yang and Bo Ai are with the School of Electronics and Information Engineering, Beijing Jiaotong University, Beijing 100044, China (e-mail:  myang@bjtu.edu.cn, boai@bjtu.edu.cn).}
}

\maketitle
\begin{abstract}
This paper investigates artificial intelligence (AI) empowered schemes for reconfigurable intelligent surface (RIS) assisted networks from the perspective of fast implementation. We formulate a weighted sum-rate maximization problem for a multi-RIS-assisted network. To avoid huge channel estimation overhead due to activate all RISs, we propose a computer vision (CV) enabled RIS selection scheme based on a single shot multi-box detector. To realize real-time resource allocation, a deep neural network (DNN) enabled transmit design is developed to learn the optimal mapping from channel information to transmit beamformers and phase shift matrix. Numerical results illustrate that the CV module is able to select of RIS with the best propagation condition. The well-trained DNN achieves similar sum-rate performance to the existing alternative optimization method but with much smaller inference time.
\end{abstract}

\section{Introduction}



Providing high-quality wireless coverage for mobile intelligent applications has become the central consideration of B5G wireless communication systems. As a wireless coverage enabler, reconfigurable intelligent surface (RIS) draws great attention due to its low energy consumption, easy deployment, and flexible programming\cite{RIS-good}. In particular, by jointly designing the transmit beamformer at the transmitter and tuning the reflecting elements on the RIS, the propagation condition is improved by mitigating the impact of obstacles on the radio frequency (RF) signals. Nevertheless, the RIS complicates the channel estimation and resource allocation, and the complexity is proportional to the number of reflecting elements, which challenges the large-scale deployment of RIS\cite{RIS-over}.


Recently, there has been increasing interest in leveraging artificial intelligence (AI) to empower wireless networks \cite{AI}. On the one hand, the computer vision (CV) methods have been employed to distinguish line-of-sight (LoS) link and Non-LoS (NLoS) link to facilitate precise channel estimation \cite{cv-channel}. On the other hand, the ``learning-to-optimize" paradigm shows great advantages in realizing near-optimal and real-time computation when handling the resource allocation problems in wireless networks \cite{dl}. Consequently, applying AI methods to tackle the challenges in designing RIS-assisted networks is a promising approach.







Thus far, there have been some existing works on AI-empowered RIS-assisted networks. In  \cite{cv-ris},  a CV-based approach to aid RIS for dynamic beam tracking was proposed, which validates the effectiveness of using visual information to control the RIS. In \cite{cv-ris2}, a CV module 
was utilized to compute the distances between RISs and mobile users for reconfigurable radiation pattern selection, and the robustness of CV against the blockages was demonstrated numerically. Nevertheless, the existing CV-assisted RIS networks activated all reflecting elements. Most existing works show that the performance gain tends to saturation with the number of reflecting elements \cite{ris-sta}. Therefore, activating partial reflecting elements not only mitigates the heavy overhead for channel estimation but also induces limited  performance loss. Besides, in \cite{dl3}, the deep neural network (DNN)-enabled transmit design was proposed for RIS-assisted networks to realize end-to-end mapping and joint beamforming of  transmitter and RIS. In \cite{dl4}, the DNN was employed to optimize the phase shift coefficients with regard to bit error rate for RIS-assisted cooperative communications. However, the supervised learning was adopted in \cite{dl3,dl4} which burdens the generation of the training set. In \cite{dl2}, the DNN-enabled transmit design was shown to achieve good performance via unsupervised learning. 



In this paper, we consider a multi-RIS-assisted multi-user multiple-input-multiple-output (MISO) network and formulate a weighted sum-rate maximization problem. To reduce channel estimation overhead, a CV-enabled RIS selection scheme is proposed to activate the RIS with the best propagation condition among a set of RISs. Moreover, to realize real-time optimization, a DNN-enabled transmit design trained via unsupervised learning is proposed to jointly optimize the transmit beamformers and the phase shift matrix. Numerical results validate the proposed schemes. It is shown that both the CV module and the DNN-based resource allocation are able to realize fast implementation with exceptional performance.

The rest of this paper is organized as follows. 
Section II introduces the system model. The proposed CV-enabled RIS selection and  DNN-enabled transmit design are presented in Section III and Section IV, respectively. Numerical results are presented in Sections V. Finally, Section VI concludes this paper.

\section{System Model}


\begin{figure*}[t]
\begin{center}
\centerline{\includegraphics[ width=1\textwidth]{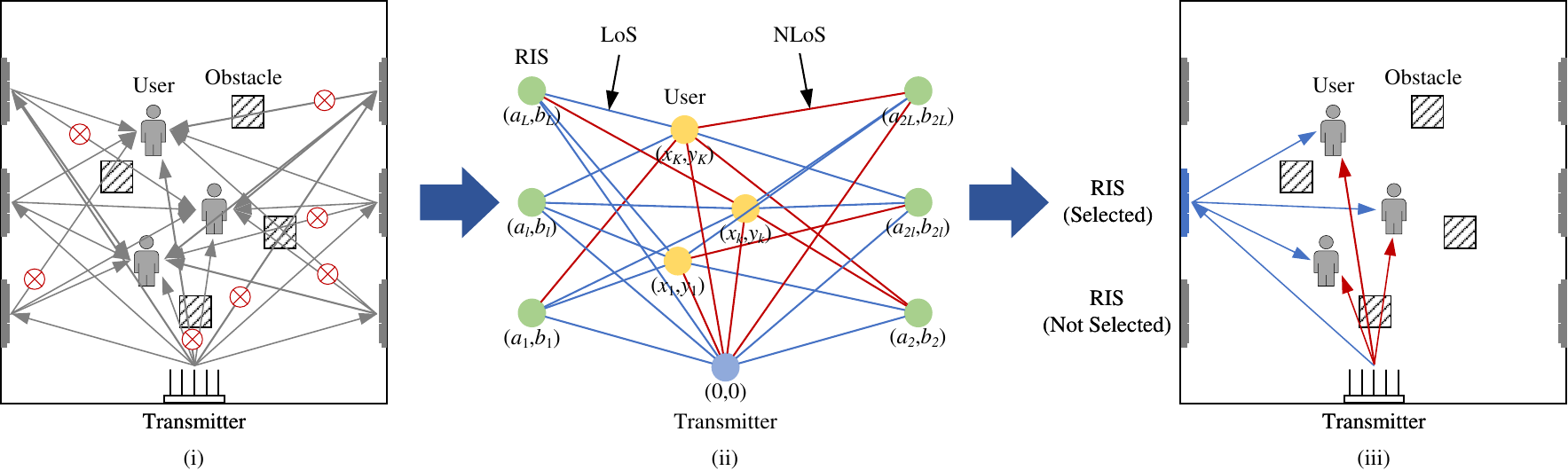}}
\caption{(i) A top view of a multi-RIS-assisted network. (ii) Graph representation via CV. (iii) RIS selection.}
\label{sys}
\end{center}
\end{figure*}

Considered a multi-RIS-assisted multi-user MISO network as shown in Figure \ref{sys}(i) where a $N_{\rm T}$-antenna transmitter serves $K$ mobile users with the help of $2L$ passive RISs. Each RIS is with $N$ reflecting elements. The transmitter intends to activate one\footnote{As each RIS is equipped with a large number of passive reflecting elements, a huge overhead for channel estimation of all transmitter-RIS links shall be induced. Therefore, we activate only one RIS at a time.} RIS of $2L$ RISs to enhance the information transmission. Due to obstacles, the LoS links may be unavailable between the transmitter/RISs and users, and propagation attenuation is induced on NLoS links. It is difficult for traditional channel estimation methods to distinguish LoS link and NLoS link in complex wireless environment \cite{channel}, which however, can be easily realized by CV. By input a top view of the considered system into a CV module, it is capability of detecting and tracking the users and obstacles.  For clarity, we use ${\cal K}\triangleq\{1,...,K\}$, ${\cal L}\triangleq\{1,...,2L\}$ and ${\cal N}\triangleq\{1,...,N\}$ to denote the sets of users, RISs and reflecting elements of each RIS, respectively. 

Denote the channel between the transmitter and the  RIS $l$ by ${\bf H}_l\in{\mathbb C}^{{N}\times{N_{\rm T}}}$, the channel between the transmitter and the user $k$ by ${\bf g}_k\in{\mathbb C}^{N_{\rm T}}$ and the channel between the RIS $l$ and the user $k$ by ${\bf h}_{l,k}\in{\mathbb C}^{N}$. The received signal at the user $k$ when the RIS $l$ is activated is given by
\begin{flalign}\label{rec_sig}
{y_{l,k}}= \left( {{\bf{g}}_k^H+{\bf{h}}_{l,k}^H{\bm\Phi}_l{{\bf{H}}}_l} \right)\sum\nolimits_{i \in \mathcal{K}} {{{\bf{w}}_i}} {s_i} + {n_k},
\end{flalign}
where $s_i$ denotes the transmit signal for user $i$ with ${\mathbb E}\{|s_i|^2\}=1$ and ${\bf w}_i$ denotes the corresponding beamformer, {${\bm\Phi}_l \triangleq {\rm diag}\{[e^{j\phi^{(l)}_{1}},e^{j\phi^{(l)}_{2}},...,e^{j\phi^{(l)}_{N}}]\} \in \mathbb{C}^{N \times N}$  denotes the phase shift matrix of the RIS $l$ with $\phi^{(l)}_{n} \in [0,2\pi)$, and $n_k\sim\mathcal{CN}( {0,{\sigma_k ^2}})$ denotes the additive white Gaussian noises (AWGN) at the 
 user $k$.

When the  RIS $l$ is activated, the received information rate at the user $k$ is given by
\begin{flalign}\label{r_u}
&{{R_k}\left( {\left\{ {{\bf{w}}_i}\right\},{{\bm\Phi} _l} } \right)} = \\
&{\log _2}\left( {1 + \frac{{{{\left| {\left( {{\bf{g}}_k^H+{\bf{h}}_k^H{\bm\Phi}_l {\bf{H}}_l} \right){{\bf{w}}_k}} \right|}^2}}}{{\sum\nolimits_{i\in\mathcal{K}\backslash k}{{{\left| {\left( {{\bf{g}}_k^H + {\bf{h}}_k^H{\bm\Phi}_l {\bf{H}}_l} \right){{\bf{w}}_i}} \right|}^2}}+ \sigma_k^2}}} \right)\nonumber
\end{flalign}

Our goal is to select the RIS with the best cascade channel condition, and then, maximize the weighted sum rate under constraints of the power budget at the transmitter and the phase shift coefficient of each reflecting element. The corresponding optimization problem is mathematically expressed as
\begin{subequations}\label{p0} 
\begin{align}
{{\bf P}_0:} &\mathop {\max }\limits_{{\left\{ {{{\bf{w}}_i}} \right\},l,{\bm\Phi}_l} }\sum\nolimits_{k \in {\cal K}} \omega_k \mathop {\max }\limits_l \left\{ {{R_k}\left( {\left\{ {{\bf{w}}_i}\right\},{{\bm\Phi} _l} } \right)} \right\} \\
{\rm s.t.}~&\sum\nolimits_{i \in {\cal K}} {\left\| {{{\bf{w}}_i}} \right\|} _2^2 \le {P_{\rm max}}, \label{p0:b}\\
&\phi^{\left(l\right)}_{n} \in \left[0,2\pi\right),\label{p0:c} \\
&\forall i,k \in {\mathcal K}, \forall l\in \mathcal{L}, \forall n \in {\mathcal N},
\end{align}
\end{subequations}
where $\omega_k$ denotes the weight factor of the  user $k$ and $P_{\rm max}$ denotes the power budget at the transmitter. To achieve fast implementation, a CV-enabled RIS selection and DNN-enabled transmit design are respectively given in the subsequent two sections. 

\section{CV-enabled RIS Selection}

A Cartesian coordinate system is defined with the location of transmitter being fixed by $(0,0)$ and that of the  RIS $l$ being fixed by $(a_l,b_l)$. Then, the effective pathloss of the $l$-th transmitter-RIS link is denoted by $$e^{({\rm T-R})}_l\triangleq{\alpha _l}{\sqrt {a_l^2 + b_l^2}},$$ where $\alpha_l\in[1,\infty)$ represents the attenuation factor, where $\alpha_l=1$ denotes the LoS link, and $\alpha_l>1$ denotes the NLoS link and the value of $\alpha_l$ is determined by the impact of the obstacles on the NLoS link. 

Denote the coordinate of the user $k$ by $(x_k,y_k)$. Then, the effective pathloss of the $(l,k)$-th RIS-user link is given by $$e^{({\rm R-U})}_{l,k}\triangleq{\beta_{l,k}}\sqrt {{{({a_l} - {x_k})}^2} + {{({b_l} - {y_k})}^2}},$$ where $\beta_{l,k}\in[1,\infty)$ represents the obstacle-aware attenuation factor. 

The RIS selection problem is expressed as
\begin{flalign}\label{RIS_s}
{l^\star} = \mathop {\arg \min }\limits_{l \in {\cal L}} \sum\nolimits_{k \in {\cal K}} \underbrace {e_l^{({\rm{T - R}})}e_{l,k}^{({\rm{R - U}})}}_{{\rm{Effective~pathloss~of~cascade~channel}}},
\end{flalign}
where ${l^\star}$ denotes the index of an RIS to be activated. 

The main idea of the CV-enabled RIS selection is to utilize the visual  information via a CV module to compute the effective pathloss of cascade channel (defined in \eqref{RIS_s}) associated with each RIS to select the RIS with the best propagation condition. Therefore, the key step to solve the problem \eqref{RIS_s} is to obtain $\{(x_k,y_k)\}$ and $\{\alpha_l,\beta_{l,k}\}$, which is realized via a single shot multi-box detector (SSD) illustrated in Figure \ref{ssd} and described as follows.

\subsubsection{SSD-based user and obstacle detection} 



With a top view of the considered system, the users and obstacles are detected by a one-stage detector in order to realize fast detection. 

In particular, the VGG16 is adopted as the backbone network to accept a frame of the top view and yield its feature maps, with which a SSD \cite{ssd} directly predicts category (i.e., user or obstacle) scores of all objects required to detect  and offsets of multiple-scale default boxes. The category  of one object is determined by comparing its category score and  a pre-defined threshold (e.g., $0.5$). Besides, the offsets of default boxes can be utilized to compute the central coordinate, height and width of the bounding box of each object.

\subsubsection{{User and obstacle location}} With the location information of bounding boxes of objects, more precise coordinates of users and obstacles can be obtained by the edge detection.

In particular, the location information of bounding boxes enables each object to be roughly cropped from the frame  to reduce the background noise during the edge detection. Then, the edge information of each object is detected via the Canny algorithm \cite{Canny} and shaped into a polygon via the polygon approximation algorithm \cite{polygon}. The coordinates of vertexes of polygon is also obtained, which are used to calculate the coordinates of users, i.e., $(x_k,y_k)$, or the coordinates of vertexes of  obstacles.

With the location information of users and  obstacles, the penetration distance of each transmitter-RIS  or RIS-user link over  obstacles is obtained, which can be utilized to calculate the corresponding attenuation factor, i.e., $\alpha_l$ or $\beta_{l,k}$.

Once $\{(x_k,y_k)\}$ and $\{\alpha_l,\beta_{l,k}\}$ are available, the problem \eqref{RIS_s} can be directly solved and the RIS with the best  propagation condition is selected with its index being the optimal solution to the problem \eqref{RIS_s}.

\begin{figure*}[t]
\begin{center}
\centerline{\includegraphics[ width=1\textwidth]{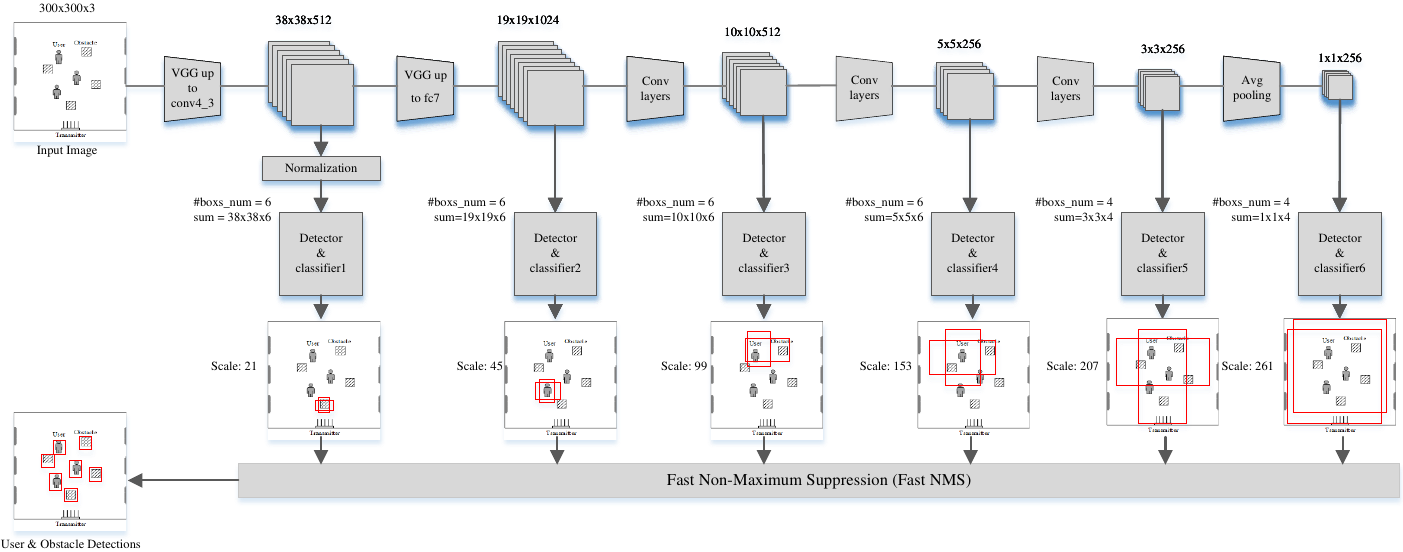}}
\caption{SSD-based user and obstacle detection in the CV module.}
\label{ssd}
\end{center}
\end{figure*}

\section{DNN-enabled Transmit Design}

Once \eqref{RIS_s} is solved, ${l^\star}$ is obtained, with which the problem \eqref{p0} is rewritten as
\begin{subequations}\label{p1}
\begin{align}
{{\bf P}_1:} &\mathop {\max }\limits_{{\left\{ {{{\bf{w}}_i}} \right\},{\bm\Phi}_{{l^\star}}} }\sum\nolimits_{k \in {\cal K}} \alpha_k {{R_k}\left( {\left\{ {{\bf{w}}_i}\right\},{{\bm\Phi} _{l^{\star}}} } \right)}  \\
{\rm s.t.}~&\phi^{\left({l^\star}\right)}_{n} \in \left[0,2\pi\right),\label{p1:b} \\
&{\rm (\ref{p0:b})}, \forall i,k \in {\mathcal K}, \forall n \in {\mathcal N}.
\end{align}
\end{subequations}
For notation simplicity, we use $\phi_{n}$ instead of $\phi _n^{\left( {{l^{ \star }}} \right)}$ and ${\bf h}_k$ instead of ${\bf h}_{{l^{ \star }},k}$.

\begin{figure}[t]
\begin{center}
\centerline{\includegraphics[ width=0.47\textwidth]{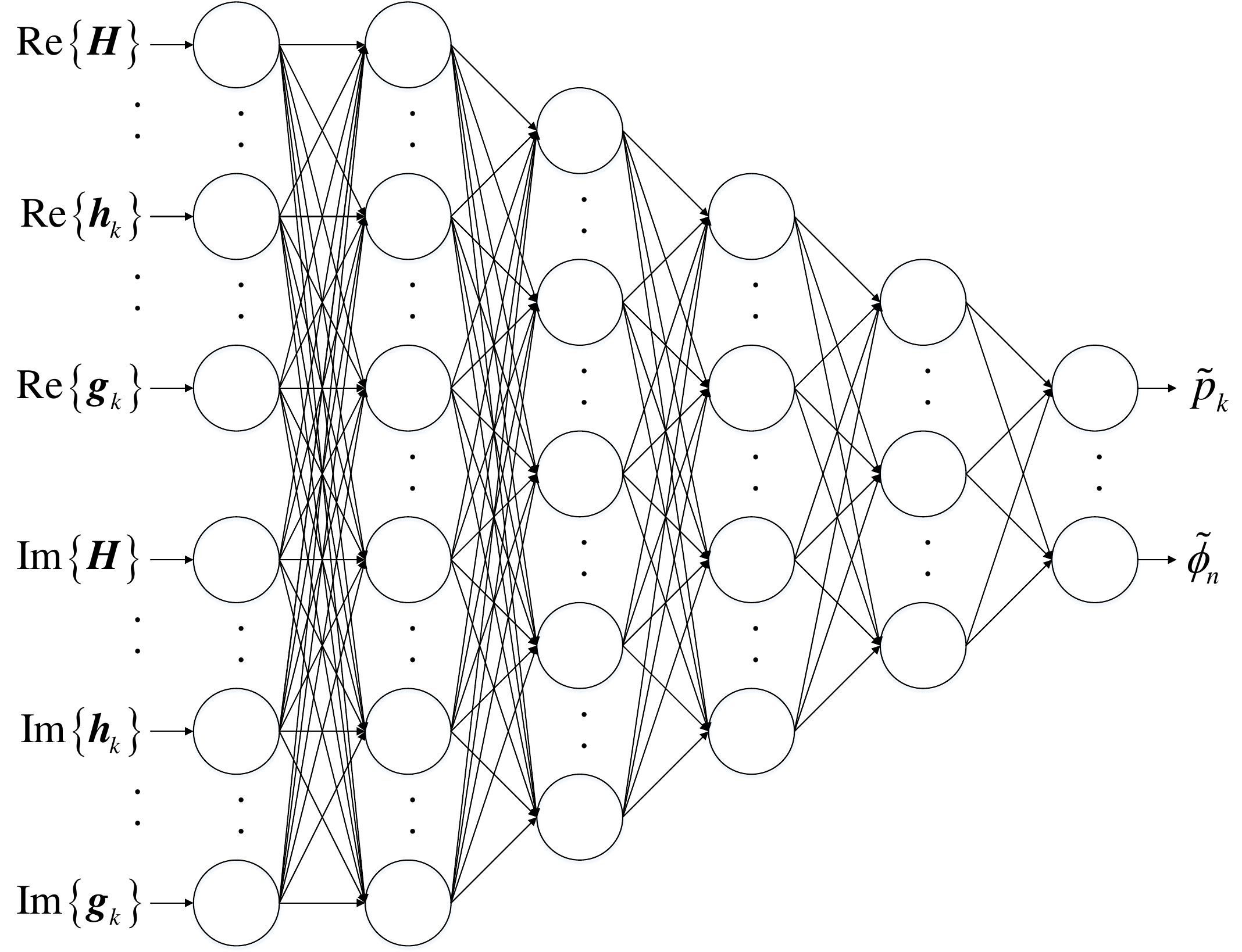}}
\caption{DNN-enabled transmit design.}
\label{dl}
\end{center}
\end{figure}

Each beamformer can be represented by
\begin{flalign}\label{beam}
{\bf w}_k={\sqrt p_k}{{\overline{\bf w}}_k},
\end{flalign}
where $p_k\in{\mathbb R}^+$ and ${{\overline{\bf w}}_k}$ satisfying $\|{{\overline{\bf w}}_k}\|_2=1$ denote the power component and the direction component, respectively.

Define
\begin{flalign}
{\bm\phi} & \triangleq {\left[ {\phi _1,\phi _2,...,\phi _N} \right]^T} \in {\mathbb R}^{N\times 1},\\
{\bf f}_k\left({\bm\phi}\right) & \triangleq {{{ { {{\bf{g}}_k^H+{\bf{h}}_k^H{\bm\Phi}_{l} {\bf{H}}_{l}} }}}} \in {\mathbb C}^{1 \times N_{\rm T}},
\end{flalign}
and
\begin{flalign}
{\bf G}\left({\bm\phi}\right)\triangleq\left[{\bf f}_1\left({\bm\phi}\right);{\bf f}_2\left({\bm\phi}\right);...;{\bf f}_K\left({\bm\phi}\right)\right] \in {\mathbb C}^{K \times {N_{\rm T}}}.
\end{flalign}

With the minimum mean square error (MMSE) scheme, the direction of the $k$-th beamformer is set by $${{\overline{\bf w}}_k} = \frac{{\bf v}_k({\bm\phi})}{\|{\bf v}_k({\bm\phi})\|_2},$$ where ${{\bf v}_k}({\bm\phi})$ is the $k$-th column of $${\bf V}({\bm\phi})\triangleq {\bf G}({\bm\phi})^H({\bf G}({\bm\phi}){\bf G}({\bm\phi})^H + \sigma^2{\bf I}_{K}  )^{-1}.$$ Then, the problem \eqref{p1} is approximated by the following the problem \eqref{p2}, i.e.,
\begin{subequations}\label{p2}
\begin{align}
{{\bf P}_2:} &\mathop {\max }\limits_{\left\{p_k,\phi_n \right\} }\sum\nolimits_{k \in {\cal K}} \alpha_k {{{\widetilde R}_k}\left( \left\{p_k,\phi _n \right\} \right)} \\
{\rm s.t.}~&\phi_{n} \in \left[0,2\pi\right),\label{p2:a} \\
&\sum\nolimits_{i \in {\cal K}} p_i \le {P_{\rm max}},\label{p2:b}\\
& \forall i,k \in {\mathcal K}, \forall n \in {\mathcal N},
\end{align}
\end{subequations}
where
\begin{flalign}
{{\widetilde R}_k}&\left( \left\{p_k,\phi _n\right\} \right) \\
&= {\log _2}\left( {1 + \frac{{{p_k}{{\left| {{{\bf{f}}_k}\left( {\bm\phi}  \right){{\overline{\bf w}}_k}} \right|}^2}}}{\sum\nolimits_{i \in {{\cal K}\backslash k}}{{p_i}{{\left| {{{\bf{f}}_k}\left( {\bm\phi}  \right){{\overline{\bf w}}_k}} \right|}^2} + \sigma _k^2}}} \right).\nonumber
\end{flalign}


Our goal is to train a DNN to learn the mapping from $\{{\bf H},{\bf h}_k,{\bf g}_k\}$ to the near-optimal solution to the problem \eqref{p2} as shown in Figure \ref{dl}. Particularly, a fully connected neural network is constructed which includes one input layer, $G$ hidden layers and one output layer. 

In particular, the input layer accepts an real-value input with the size of 
\begin{flalign}\label{a_1}
a_1 \triangleq 2\left(N\times N_{\rm T}+K\times \left(N + N_{\rm T}\right)\right).
\end{flalign}
 For the $g$-th ($l\in\{1,2,...,G\}$) hidden layer, denote the learnable weights by ${\bm\Omega}_g\in {\mathbb R}^{{a_{g+1}}\times{a_g}}$ and ${\bm\xi}_g\in {\mathbb R}^{{a_{g+1}}}$, where $a_g/a_{g+1}$ represents the input/output dimension of the $g$-th hidden layer. Then, the output of the $g$-th hidden layer is given by
\begin{flalign}
 {\bf d}_{g+1} = {\rm ReLU}\left({\bm\Omega}_g {\bf d}_g + {\bm\xi}_g \right),~g\in\{1,2,...,G\},
\end{flalign}
where ${\bf d}_g \in {\mathbb R}^{a_g}$/${\bf d}_{g+1} \in {\mathbb R}^{a_{g+1}}$ represents the input/output of the $g$-th hidden layer and ${\rm ReLU}(\cdot)$ represents the ReLU activation function for nonlinear mapping. The output layer is to shape ${\bf d}_{G+1}$ with the same dimension of $\{p_k,\phi_n\}$, i.e., $(K+N)$,
which is given by
\begin{flalign}
\left[{\widetilde p}_1, {\widetilde p}_2,...,{\widetilde p}_K \right]^T &= {\rm Sig}\left( {\bm\Theta}_1{\bf d}_{G+1} + {\bm \mu }_1 \right),\\
\left[{\widetilde \phi}_1, {\widetilde \phi}_2,...,{\widetilde \phi}_N \right]^T &= {\rm Sig}\left( {\bm\Theta}_2{\bf d}_{G+1} + {\bm \mu }_2 \right),
\end{flalign}
where ${\bm\Theta}_1\in{\mathbb R}^{K \times a_{G+1} }$, ${\bm\Theta}_2\in{\mathbb R}^{N \times a_{G+1}}$, ${\bm \mu }_1\in{\mathbb R}^{K}$ and ${\bm \mu }_2\in{\mathbb R}^{N}$  denote the learnable weights of the output layer, and ${\rm Sig}(\cdot)$ represents the Sigmoid activation function. Further, to guarantee the output of the DNN satisfies \eqref{p2:a} and \eqref{p2:b}, $\{{\widetilde p}_k,\widetilde{\phi}_n\}$ are respectively updated\footnote{As the problem \eqref{p2} aims to maximize the sum rates, the transmitter intends to use all power budget, and \eqref{pk_update} guarantees that $\sum\nolimits_{i \in {\cal K}} {\widetilde{p}}_i = {P_{\rm max}}$. } by
\begin{flalign}
&{\widetilde p}_k := \frac{{{e^{{{\widetilde p}_k}}}}}{{\sum\nolimits_{i \in {\cal K}} {{e^{{{\widetilde p}_i}}}} }}{P_{{\rm{max}}}},~\forall k\in {\mathcal K},\label{pk_update}\\
&\widetilde{\phi}_n := 2\pi \widetilde{\phi}_n,~\forall n\in {\mathcal N}. 
\end{flalign}

Additionally, to deal with the disappearance of the gradient due to the Sigmoid function, the batch normalization operation is adopted to process the output data after each hidden layer.

Note that the proposed DNN is able to generate a feasible solution to the problem \eqref{p2}  and thus, we can train the DNN to be competent to solve problem \eqref{p2}  via unsupervised learning. Define ${\bm \theta}\triangleq \{{\bm\Omega}_g,{\bm\xi}_g,{\bm\Theta}_g,{\bm\Theta}_2,{\bm \mu }_1,{\bm \mu }_2\}$ to represent all the learnable weights of the DNN. With a batch of $M$ training samples, i.e., $\{{\bf H}^{(m)},{\bf h}_k^{(m)},{\bf g}_k^{(m)}\}_{m=1}^{M}$, the corresponding outputs with ${\bm \theta}$ are given by $\{{\widetilde p}_k^{(m)},{\widetilde \phi}_n^{(m)}| {\bm \theta} \}_{m=1}^{M}$. Then, the loss function  is given by
\begin{flalign}\label{Loss_Function}
{\cal L}_M\left( {\bm \theta}  \right) = -\frac{1}{M}\sum\nolimits_{k \in {\cal K}} \alpha_k {{R_k}\left( \left\{{\widetilde p}_k^{(m)},{\widetilde \phi}_n^{(m)}| {\bm \theta} \right\} \right)}
\end{flalign}
with which ${\bm \theta}$ is updated via the Adam method by
\begin{equation}\label{up_theta}
{\bm \theta}  := {\bm \theta } - \frac{\omega {{{\bm v}_t}\left(\nabla {\cal L}_M\left( {\bm \theta}  \right)\right)}}{\sqrt{{{\bm s}_t}\left(\nabla {\cal L}_M\left( {\bm \theta}  \right)\right) + \varepsilon}},
\end{equation}
where $\omega$ denotes the learning rate, $\varepsilon$ is a hyperparameter, ${{{\bm v}_t}(\nabla {\cal L}_M( {\bm \theta}  ))}$ and ${{\bm s}_t}(\nabla {\cal L}_M( {\bm \theta}  ))$ are two functions defined by Adam to modify gradient and learning rate, respectively.

\begin{algorithm}[t]
\caption{Training DNN for solving problem \eqref{p2}}
 {\bf Training set:}  $\{{\bf H}^{(d)},{\bf h}_k^{(d)},{\bf g}_k^{(d)}\}_d$\;
 {\bf Initialize}  $\bm\theta$\;
 \For{{\rm epoch} $e \in [0,1,\dots,E] $}{
 \For{{\rm minibatch} $b \in [0,1,\dots,B] $}{Sample $\{{\bf H}^{(m)},{\bf h}_k^{(m)},{\bf g}_k^{(m)}\}_{m=1}^{M}$ from the training set\;
 Obtain the output of DNN $\{{\widetilde p}_k^{(m)},{\widetilde \phi}_n^{(m)}| {\bm \theta} \}_{m=1}^{M}$\;
 Calculate the loss function ${\cal L}_M\left( {\bm \theta}  \right)$ via (\ref{Loss_Function})\;
 Update ${\bm \theta}$ via (\ref{up_theta})\;
 }
}
{\bf Return}  $\bm\theta$.
\label{UL}
\end{algorithm}

The training process of DNN is summarized in Algorithm 1. With the output of DNN, i.e., $\{{\widetilde p}_k,\widetilde{\phi}_n\}$, the beamformers can be recovered via \eqref{beam}. Besides, the computational complexity of DNN-enabled transmit design is on the order of {${\mathcal O}({  \sum\nolimits_{g=1}^{G} {({a_{g+1}}{a_g^2} + {a_{g+1}}}) +(K+N)(1+{a_{G+1}^2}) })$}, where the problem dimension is relevant to $a_1$ defined in \eqref{a_1}.

\section{Numerical Results}

This section provides numerical results to evaluate the CV-enabled RIS selection and DNN-enabled transmit design.

{\bf Simulation scenario:} The numbers of RISs, reflecting elements of each RIS, transmit antennas and mobile users are respectively set by $2L=6$, $N = 8$, $N_{\rm T} = 8$, and $K = 4$. The power budget of the transmitter is set by $P_{\rm max} = 1$, the average ratio between path loss and noise power is set to $20$ dB. 


{\bf Datasets:} 
The CV module is trained on a dataset of top views of the considered system. In each sample, the coordinates of users and obstacles as well as the number of  obstacles are randomly generated. 

The DNN is trained on a set of  $100,000$ unlabeled samples. Besides, the test set includes $1,000$ labeled samples and each sample has two labels, i.e., the result of $25$ iterations via the AO-based algorithm (denoted by AO 25) and the result of $50$ iterations via the AO-based algorithm (denoted by AO 50).


Our implementation is developed using Python 3.9.18 with Pytorch 2.2.0 on a computer with Intel(R) Core(TM)  i5-9300H CPU and one NVIDIA RTX 1650 GPU (4 GB of memory).





\subsubsection{Test performance of CV module} Figure \ref{ris_s} illustrates the RIS selection by the CV module. It is observed that the selected RIS suffers from the least penetrate loss and multiplicative fading, which validates the effectiveness of the CV-enabled RIS selection. Besides, the CV module achieves $11$ FPS for $300\times 300$ input (i.e., the top view). Note that a CV selection result is applicable for several frames due to the low-speed mobility of users and obstacles. 

\begin{figure}[t]
\begin{center}
\centerline{\includegraphics[ width=0.49\textwidth]{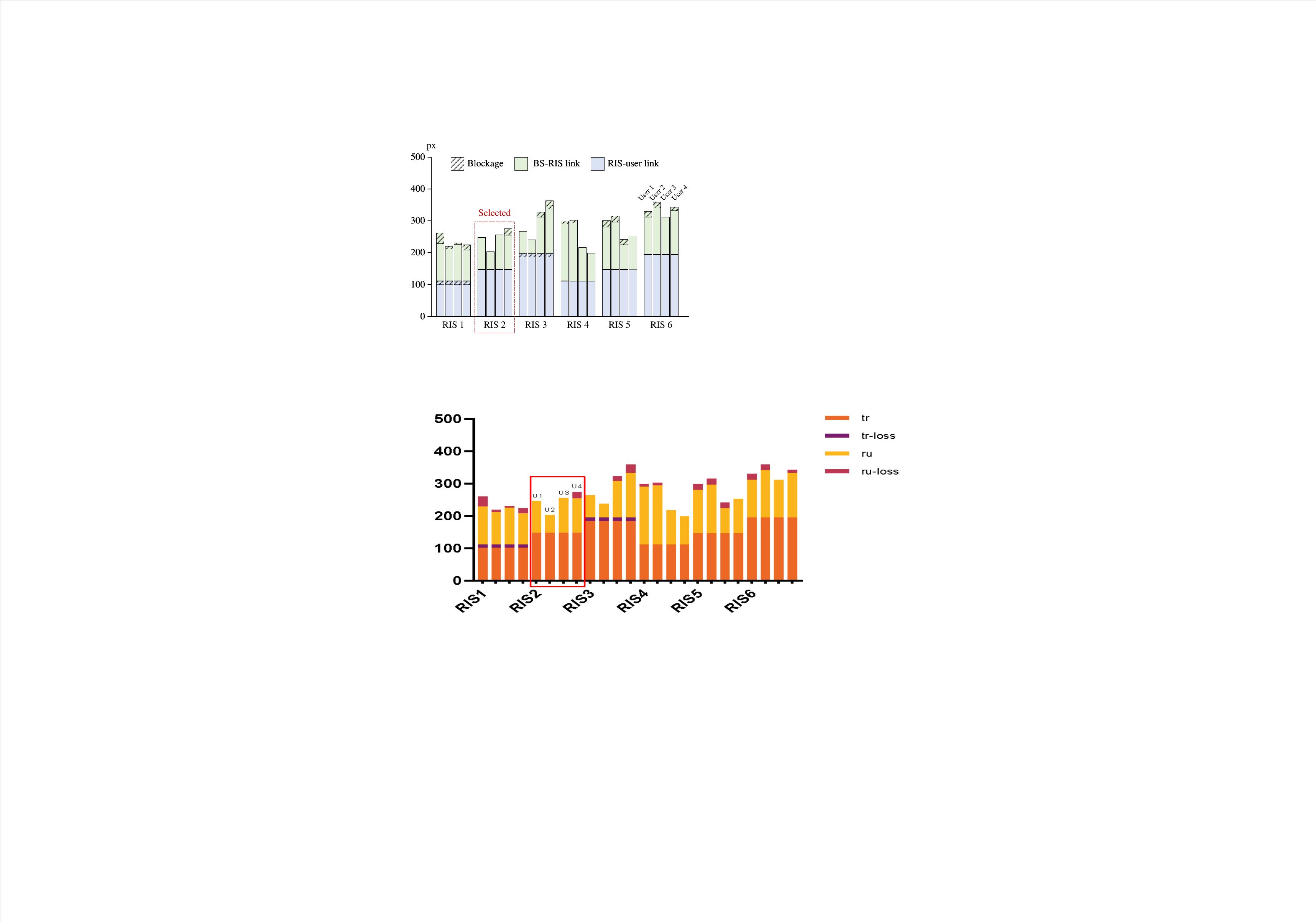}}
\caption{Illustration of the CV-enabled RIS selection.}
\label{ris_s}
\end{center}
\end{figure}



\subsubsection{Test performance of DNN}


In Table \ref{table:3}, the average sum rate over $1,000$ test samples yield by the well-trained DNN is very close to that by the AO-based algorithm (i.e., the performance loss is lower than $10\%$) while the inference speed of DNN is one million faster than the AO-based algorithm which is able to realize real-time optimization. 

\begin{table}[h]
\centering
\caption{Evaluation of DNN-enabled transmit design based on average sum rate and average inference time. }
\begin{tabular}{c|| c|c}
\hline
Method &Sum rate [bit/s/Hz] & Inference time  \\
 \hline
  \hline
    AO 25 & $37.83$ & $69.8$ s\\
 \hline
    AO 50 & $38.02$ & $118.9$ s\\
 \hline
 DNN & $34.44$ &{$0.02$ ms}\\
 \hline
\end{tabular}
\label{table:3}
\end{table}

\section{Conclusion}
This paper applied two AI methods to design multi-RIS-assisted networks to realize fast implementation. A CV module was designed to select the RIS with the best RF signal propagation condition among a set of RISs to reduce the channel estimation overhead. With the selected RIS, a DNN was trained to jointly optimize the transmit beamformers and phase shift matrix. The result of the CV-enabled RIS selection scheme was illustrated. The well-trained DNN  was shown to be close to the AO-based algorithm but with much faster inference speed. These results demonstrate the advantage of applying AI to wireless networks.

\end{document}